\documentclass[12pt]{article}
\usepackage[reqno]{amsmath}
\usepackage{epsfig}
\usepackage{array}
\usepackage{float}
\usepackage{bbm}

\usepackage{a4}
\usepackage{epsfig}
\usepackage{a4wide}
\usepackage{wasysym}

\def\ra{\rightarrow}
\def\be{\begin{equation}}
\def\ee{\end{equation}}
\def\gs{\mathrel{
   \rlap{\raise 0.511ex \hbox{$>$}}{\lower 0.511ex \hbox{$\sim$}}}}
\def\ls{\mathrel{
   \rlap{\raise 0.511ex \hbox{$<$}}{\lower 0.511ex \hbox{$\sim$}}}}

\newcommand{\onbb}{neutrinoless double beta decay}

\newcommand{\ba}{\begin{array}{c}}
\newcommand{\baz}{\begin{array}{cc}}
\newcommand{\bad}{\begin{array}{ccc}}
\newcommand{\bav}{\begin{array}{cccc}}
\newcommand{\bea}{\begin{equation} \begin{array}{c}}
\newcommand{\eea}{ \end{array} \end{equation}}
\newcommand{\ea}{\end{array}}
\newcommand{\D}{\displaystyle}
\newcommand{\dms}{\mbox{$\Delta m^2_{\odot}$}}
\newcommand{\dma}{\mbox{$\Delta m^2_{\rm A}$}}
\newcommand{\meff}{\mbox{$\left| < \! m \! > \right|$}}

\def\ra{\rightarrow}

\def\gtap{\mathrel{
   \rlap{\raise 0.511ex \hbox{$>$}}{\lower 0.511ex \hbox{$\sim$}}}}
\def\ltap{\mathrel{
   \rlap{\raise 0.511ex \hbox{$<$}}{\lower 0.511ex \hbox{$\sim$}}}}

\newcommand{\betabeta}{\mbox{$(\beta \beta)_{0 \nu}  $}}

\newcommand{\pmns}{\mbox{$ U_{\rm PMNS}$}}

\newcommand{\ts}{\mbox{$ \tan^2 \theta_{\rm \odot}$}}
\newcommand{\sa}{\mbox{$ \sin^2 2\theta_{\rm atm}$}}

\begin{document}

\title{
\vspace{-2cm}
\hfill {\small SISSA 63/2004/EP}\\
\vspace{-0.3cm}
\hfill {\small hep-ph/0409135} \\ 
\vskip 0.8cm
\bf Flavor Symmetry $L_e - L_\mu - L_\tau$, Atmospheric
Neutrino Mixing and CP Violation in the Lepton Sector
}

\author{
S.T.~Petcov\thanks{Also at: Institute of 
Nuclear Research and Nuclear Energy,
Bulgarian Academy of Sciences, 1784 Sofia, Bulgaria}~
$\;$ and $\;$ W.~Rodejohann\\ \\
{\normalsize \it Scuola Internazionale Superiore di Studi Avanzati}\\
{\normalsize \it Via Beirut 2--4, I-34014 Trieste, Italy}\\
{\normalsize and}\\
{\normalsize \it Istituto Nazionale di Fisica Nucleare}\\
{\normalsize \it Sezione di Trieste, I-34014 Trieste, Italy}
}
\date{}
\maketitle
\thispagestyle{empty}
\vspace{-0.8cm}
\begin{abstract}
\noindent The PMNS neutrino mixing matrix is given, in general, 
by the product of two unitary matrices 
associated with the diagonalization of the charged lepton 
and neutrino mass matrices.
Assuming that the active flavor 
neutrinos possess a 
Majorana mass matrix which is 
diagonalized by a 
bimaximal mixing matrix, we 
give the allowed forms of the 
charged lepton mixing matrix and 
the corresponding implied forms of the 
charged lepton mass matrix. 
We then assume that the 
origin of bimaximal mixing is a weakly  
broken flavor symmetry  
corresponding to the conservation of the non--standard lepton charge
$L' = L_e - L_\mu - L_\tau$.   
The latter does not predict, in general, the 
atmospheric neutrino mixing to be maximal. 
We study the impact of 
this fact on the allowed forms of the charged lepton 
mixing matrix and on the 
neutrino mixing observables, analyzing the 
case of $CP$--violation in detail. 
When compared with the case of exact 
bimaximal mixing, the deviations 
from zero $U_{e3}$ and from maximal atmospheric neutrino mixing 
are typically more sizable if 
one assumes just $L'$ conservation. 
In fact, $|U_{e3}|^2$ can be as small 
as 0.007 and atmospheric neutrino mixing 
can take any value inside its currently allowed range. 
We discuss under which conditions the atmospheric neutrino mixing angle 
is larger or smaller than $\pi/4$. 
We present also a simple see--saw realization 
of the implied light neutrino Majorana 
mass matrix and consider leptogenesis in this
scenario.

\end{abstract}

\newpage
\section{\label{sec:intro}Introduction}
\vspace{-0.2cm}

 There has been a remarkable progress in the studies of neutrino
oscillations in the last several years.
The experiments with solar, 
atmospheric and reactor neutrinos 
\cite{sol,SKsolar,SNO,SNO3,SKatm,SKatm04,KamLAND,KamLANDnew} 
have provided compelling evidences for the 
existence of neutrino oscillations 
driven by non--zero neutrino masses and neutrino mixing.
Evidences for oscillations of neutrinos were
obtained also in the first long baseline
accelerator neutrino experiment K2K \cite{K2K}.

 The interpretation of the solar and
atmospheric neutrino, and of K2K and KamLAND
data in terms of 
neutrino oscillations requires
the existence of 3--neutrino mixing
in the weak charged lepton current: 
\begin{equation}
\nu_{l \mathrm{L}}  = \sum_{j=1}^{3} U_{l j} \, \nu_{j \mathrm{L}}~.
\label{3numix}
\end{equation}
\noindent Here $\nu_{lL}$, $l  = e,\mu,\tau$,
are the three left--handed flavor 
neutrino fields,
$\nu_{j \mathrm{L}}$ is the 
left--handed field of the 
neutrino $\nu_j$ having a mass $m_j$
and $U$ is the Pontecorvo--Maki--Nakagawa--Sakata (PMNS) 
neutrino mixing matrix \cite{PMNS}. 
Actually, all existing neutrino 
oscillation data, except
the data of the LSND experiment \cite{LSND}~ 
\footnote{In the
LSND experiment indications 
for oscillations 
$\bar \nu_{\mu}\to\bar \nu_{e}$  
with $(\Delta m^{2})_{\rm{LSND}}\simeq 
1~\rm{eV}^{2}$ were obtained. 
The  LSND results are being tested 
in the MiniBooNE experiment
at Fermilab \cite{MiniB}.},
can be described if we assume the existence of
3--neutrino mixing in vacuum, Eq.\ (\ref{3numix}),
and we will consider this possibility in what follows.

  In the standardly used parametrization, 
the PMNS mixing matrix has the form:
\bea \label{eq:Upara}
\pmns = \left( \bad 
c_{12} c_{13} & s_{12} c_{13} & s_{13}  \\[0.2cm] 
-s_{12} c_{23} - c_{12} s_{23} s_{13} e^{i \delta} 
& c_{12} c_{23} - s_{12} s_{23} s_{13} e^{i \delta} 
& s_{23} c_{13} e^{i \delta} \\[0.2cm] 
s_{12} s_{23} - c_{12} c_{23} s_{13} e^{i \delta} & 
- c_{12} s_{23} - s_{12} c_{23} s_{13} e^{i \delta} 
& c_{23} c_{13} e^{i \delta} \\ 
               \ea   \right) 
 {\rm diag}(1, e^{i \alpha}, e^{i \beta}) \, , 
\eea
%
where we have used the usual
notations $c_{ij} = \cos\theta_{ij}$, 
$s_{ij} = \sin\theta_{ij}$,
$\delta$ is the Dirac $CP$ violation phase, 
$\alpha$ and $\beta$ are two possible Majorana 
$CP$ violation phases \cite{BHP80,Doi81}. 

 The ranges of values of
the three neutrino 
mixing angles, which are allowed 
at 3 s.d.\ by the most recent solar, 
atmospheric 
(and long--baseline accelerator) 
neutrino data and by the data 
from the reactor antineutrino experiments
CHOOZ \cite{CHOOZ}
and KamLAND, read  
\cite{SKatm04,KamLANDnew,BCGPRKL2}:
\be 
\label{eq:range}
\ba
0.27 \leq \tan^2 \theta_{\odot} \equiv 
\tan^2 \theta_{12} \leq 0.58~,\\[0.3cm]
|U_{e3}|^2 = \sin^2\theta_{13} <  0.048~, \\[0.2cm]
\sin^2 2 \theta_{\rm atm} \equiv \sin^2 2 \theta_{23} \geq 0.85~.
\ea
\ee

  The values of $\tan^2 \theta_{12}$, 
$\sin^2\theta_{13}$ and $\sin^2 2 \theta_{23}$ suggested
by the data, are relatively close to those
obtained assuming that $U_{\rm PMNS}$ has bimaximal mixing 
form:  the bimaximal mixing ``scenario'' 
\cite{bimax} would correspond to 
$\theta_{12} = \theta_{23} = \pi/4$ and $\theta_{13} = 0$.
Whereas data favors maximal atmospheric neutrino mixing, 
and allows for $\theta_{13} = 0$, 
maximal solar neutrino mixing is ruled out 
at close to 6$\sigma$ \cite{KamLANDnew,BCGPRKL2}.
Though definitely large, the atmospheric neutrino mixing angle
can deviate significantly from being maximal, 
a fact seen clearly in re--writing 
the 99.73\% (90\%) C.L.\ allowed range of
$\sin^2 2 \theta_{\rm atm} \geq 0.85~(0.92)$ as 
$0.44~(0.55)\le \tan^2\theta_{\rm atm} \le 2.26~(1.79)$.

 In what regards the neutrino mass squared 
differences driving the solar and
atmospheric neutrino oscillations, 
$\dms$ and $\dma$, the best--fits of the 
current data are obtained for
$(\dms)_{\rm BF} = 8.0 \cdot 10^{-5}$ eV$^2$ 
\cite{BCGPRKL2}
and $(\dma)_{\rm BF} = 2.1 \cdot 10^{-3}$ eV$^2$  
\cite{SKatm04}.
A phenomenologically very interesting 
quantity is the ratio $R$ of $\dms$ and $\dma$,
whose ``best--fit value'' is  
\be \label{eq:obsdm}
R_{\rm BF} \equiv \frac{(\dms)_{\rm BF}}{(\dma)_{\rm BF}} \simeq 
\frac{8.0 \cdot 10^{-5}}{2.1 \cdot 10^{-3}} \simeq 0.040 ~.
\ee 
Using for $\dma$ and $\dms$ the 3$\sigma$ allowed ranges  
from \cite{SKatm04} and \cite{BCGPRKL2},
$\dma = (1.3 - 4.2) \cdot 10^{-3}$ eV$^2$ and
$\dms = (7.2 - 9.5) \cdot 10^{-5}$ eV$^2$, 
we find that $R$ lies approximately 
between 0.017 and 0.073.  

  The understanding of the origin of the patterns 
of neutrino mixing and of neutrino mass 
squared differences suggested by the data,
is one of the central problems in today's 
neutrino physics (see, e.g., \cite{STPNu04}).
Consider the Lagrangian which is 
compatible with the low energy 
lepton phenomenology and includes a Majorana mass term for the 
left--handed active (flavor) neutrinos:
\be \label{eq:Llow} 
{\cal L} = -\frac{1}{2}~\overline{(\nu')_L}~m_\nu~(\nu')_R^c 
- \overline{(\ell')_L}~m_\ell~(\ell')_R 
+ \frac{g}{ \sqrt{2}}
W_\mu~\overline{(\ell')_L}~\gamma^\mu~(\nu')_L + h.c.~,
\ee 
where $(\ell'_{L(R)})^T \equiv (e'_{L(R)},\mu'_{L(R)},\tau'_{L(R)})$,
$(\nu'_L)^T \equiv (\nu'_{e'L},\nu'_{\mu'L},\nu'_{\tau'L})$,
$\ell'_{L(R)}$ and $\nu'_{l'L}$ being the weak eigenstate
charged lepton and neutrino fields, 
$(\nu')_{l'R}^c \equiv C \overline{(\nu')_{l'L}}^T$,
and $C$ is the charge conjugation matrix.
The neutrino mass matrix $m_\nu$ and the charged 
lepton mass matrix $m_\ell$ 
are diagonalized through 
the well--known congruent and bi--unitary 
transformations, respectively:
\be \label{eq:mnumlep}
m_\nu = U_\nu~m_\nu^{\rm diag}~U_\nu^T~,~~~  
m_\ell = U_L~m_\ell^{\rm diag}~U_R^\dagger~.
\ee
Therefore, the matrix $m_\ell m_\ell^\dagger$ is diagonalized by $U_L$. 
Written in terms of the new fields
$\nu_L = U_\nu^\dagger~(\nu')_L$, 
$\ell_L = U_L^\dagger~(\ell')_L$ and $\ell_R = U_R^\dagger~(\ell')_R$, 
the mass terms are diagonal 
and the weak charged lepton current  
takes the form:
$\overline{\ell_L}~ \gamma^\mu~U_L^\dagger U_\nu~\nu_L$. We 
therefore identify the PMNS matrix as
\be \label{eq:PMNS}
\pmns = U_L^\dagger U_\nu~.
\ee

 In the present article we extend 
the analysis performed in \cite{FPR}
on a possible origin of the patterns 
of neutrino mixing and of neutrino mass 
squared differences emerging from the data.
Following \cite{FPR}, we assume 
that $U_\nu$ has bimaximal mixing form 
and that the deviations from this form
are induced by $U_L^\dagger$. 
Such a possibility has been considered 
phenomenologically a long time ago 
(in a different context) in \cite{lelmlt}.
More recently it has been 
investigated in some detail 
also in Refs.\ \cite{old,WR,roma,andrea}. 
Certain aspects of it have 
been discussed in the context of GUT
theories, e.g., in Ref.\ \cite{lelmlt1}.
The alternative hypothesis, namely,
that  $U_L^\dagger$ has 
bimaximal mixing form
and the observed deviations
from it are due to $U_\nu$,
was considered recently 
in Refs.\ \cite{pires,SM}. The present study concentrates on the possibility 
that the non--standard lepton 
charge $L' = L_e - L_\mu - L_\tau$ is responsible for the structure of 
$U_\nu$. We note that this flavor symmetry 
does not predict the atmospheric neutrino 
mixing to be necessarily maximal, but rather to be
determined by ratio of 
two numbers of order one. This has some impact on the neutrino mixing 
observables obtained after including the corrections from the 
charged lepton sector.

In Section \ref{sec:mix} we 
review the basics of the approach
to understand the observed neutrino mixing
as deviation from bimaximal mixing. 
We then concentrate on the case 
when the charged lepton sector is 
responsible for the deviation and give 
a ``catalog'' (although not complete) 
of the implied possible 
structures of the corresponding 
charged lepton and neutrino
mass and mixing matrices. 
Considering the approximate 
conservation of the 
non--standard lepton 
charge $L' = L_e - L_\mu - L_\tau$  
by $m_{\nu}$ as a possible origin of
maximal and/or zero neutrino mixing 
we generalize the preceding 
analysis to the case of {\it a priori} 
non--maximal atmospheric neutrino 
mixing in Section \ref{sec:gen}. 
A very simple see--saw realization 
of the neutrino mass matrix 
having the indicated approximate
symmetry is given in Section \ref{sec:real}, 
where we discuss also its 
implications for leptogenesis.
Section \ref{sec:concl} summarizes the results
of this work.


\section{\label{sec:mix}Deviations from Bimaximal Neutrino Mixing}

\subsection{\label{sec:general}General considerations}


  The primary goal of the approaches 
used to produce deviations from bimaximal mixing 
is to generate a deviation from
maximal solar neutrino mixing
\footnote{It is even possible to construct 
a model in which maximal solar neutrino 
mixing would be linked to a 
vanishing baryon asymmetry of the 
Universe \cite{ichlepto}.}. 
The latter can be 
parameterized through a small parameter 
$\lambda \sim 0.2$  as \cite{WR} 
\be \label{eq:lam}
\sin \theta_{\odot} = \sqrt{\frac{1}{2}}~(1 - \lambda)~.
\ee
Note that the parameter $\lambda$ is very similar 
in value to the Cabibbo angle \cite{WR}: $\lambda \sim  \sin\theta_C \sim 
\theta_C$. 
The implied relation $\theta_\odot = \pi/4 - \theta_C$,
if confirmed experimentally, might be
linked to Grand Unified Theories \cite{martti,SM,FM}.

    When one starts from bimaximal neutrino 
mixing and introduces some  
``perturbation'' to generate 
non--maximal solar neutrino mixing, 
typically also $|U_{e3}|$ 
and the atmospheric neutrino mixing 
will deviate from their ``bimaximal'' values. 
These additional deviations will, in general, 
be proportional to some power of $\lambda$.
Thus, a useful parametrization taking 
into account this fact is \cite{WR} 
\be
U_{e3} = A_\nu~\lambda^n e^{i\beta}~,~~ 
U_{\mu 3} = \sqrt{\frac{1}{2}}~(1 - B_\nu~\lambda^m)~
e^{i (\delta + \beta)}~, 
\ee
%
where $m, n$ are integer, and $A_\nu, B_\nu$
are real numbers, $A_\nu \geq 0$.
Once $\sin \theta_{\odot}$, and correspondingly $\lambda$,
is determined with relatively high precision, 
$m, n$, $A_\nu$ and $B_\nu$ 
can be fixed using the measured values of 
$|U_{e3}|$ and $|U_{\mu 3}|$. For further
details, we refer to \cite{WR}. 

 Deviations from bimaximal neutrino mixing
can be generated, e.g., via radiative corrections 
\cite{devrad} or via small perturbations 
to a zeroth order neutrino mass matrix corresponding to 
bimaximal mixing. 
The latter can be achieved through 
``anarchical'' perturbations \cite{andre} 
or, e.g., by the presence of a small contribution 
from the conventional see--saw term in 
type II see--saw models \cite{ichII}.

   The ratio $R$ of $\dms$ and $\dma$ 
numerically turns out to be of order $\lambda^2$. Indeed, using 
the definition of $\lambda$ in Eq.\ (\ref{eq:lam}) and the range 
of $\theta_{12}$ as given in Eq.\ (\ref{eq:range}), one finds that the 
3$\sigma$ allowed range of $\lambda$ is 
\be \label{eq:lamran}
\lambda \simeq 0.16 -  0.37~. 
\ee
The best--fit value of $\sin^2\theta_{12} \simeq 0.28$, 
obtained in the 
3--neutrino oscillations analysis
in \cite{BCGPRKL2},  
corresponds to $\lambda \simeq 0.25$. 
Remembering that $R$ lies between 0.017 and 0.073, one sees that 
to a relatively good precision we have 
$R \simeq \lambda^2$. This can be used to analyze the 
structure of the neutrino mass matrix 
in a simple manner \cite{WR}. 

When the deviations from bimaximal 
mixing stem from charged lepton mixing, 
and the mass matrices are not 
implemented in a specific model, 
the ratio $R$ is independent of the deviations 
and the relation between 
$R$ and $\lambda$ is purely accidental
\footnote{As is well--known, if the 
approximate conservation of the lepton charge
$L' = L_e - L_\mu - L_\tau$ \cite{lelmlt} 
is at the origin of 
bimaximal neutrino mixing and zero \dms,
it is possible to generate 
the observed values of neutrino mass 
and mixing parameters by adding small 
perturbations to the neutrino mass matrix 
{\it only if} the effect of 
$U_L \neq {\mathbbm 1}$, arising from the 
charged lepton sector, is taken into account
\cite{Koide}, see below.}.
If alternatively
bimaximal mixing {\it and} zero $\dms$
are due to some symmetry, 
and the symmetry breaking involves 
just one ``small'' parameter, 
all deviations from bimaximal mixing
and the ratio $R$ will be linked to $\lambda$. 
An example of such a scenario 
can be found in \cite{ichII}.

  Assuming that the neutrino 
mass matrix generates bimaximal neutrino mixing
\footnote{Obviously, in the Standard Theory there 
would be no difference whether the 
bimaximal mixing originates from the
neutrino or charged lepton 
mass term. The underlying theory whose low 
energy limit is the Standard Theory 
is assumed here to have 
``chosen'' the neutrino mass term 
as the source of bimaximal mixing, 
i.e., neutrinos to be ``special''.}, 
the allowed structures of the 
charged lepton mixing matrix 
have been analyzed in a model--independent way
in \cite{FPR}. In the next Section we 
will briefly summarize 
the obtained results. 
We will give also the implied 
structures of the charged lepton mass matrix,
which were not presented in \cite{FPR}.


\subsection{\label{sec:mlep} Deviations Due to 
Charged Lepton Mixing and Implied Structures of 
the Charged Lepton Mass Matrix}


The goal is to generate successful neutrino 
mixing phenomenology by 
using the relation $\pmns = U_L^\dagger U_\nu$, 
where $U_L$ is the mixing matrix 
diagonalizing $m_\ell m_\ell^\dagger$ and 
$U_\nu$ diagonalizes the neutrino mass matrix. 
The matrix $U_\nu$ is assumed to have 
exact bimaximal mixing form, namely 
\be
U_\nu = 
\left(
\bad
\frac{1}{\sqrt{2}} & \frac{1}{\sqrt{2}} & 0 \\[0.3cm]
-\frac{1}{2} & \frac{1}{2} & \frac{1}{\sqrt{2}} \\[0.3cm]
\frac{1}{2} & -\frac{1}{2} & \frac{1}{\sqrt{2}}
\ea
\right)~. 
\ee
Three 
generic structures of $U_L$ can 
be identified in this case \cite{FPR}. 
They can be classified by the 
magnitude of the sines 
of the three Euler angles $\theta_{ij}$ in $U_L$, 
$\sin \theta_{ij} \equiv \lambda_{ij}$,
where $ij = 12, 13, 23$ in a parametrization of $U_L$  
analogous to that in Eq.\ (\ref{eq:Upara}). 
  
\begin{itemize}
\item All $\lambda_{ij}$ are small: $\lambda_{ij} \ls 0.35$.\\
The matrix $U_L$ has a hierarchical
``CKM--like'' form,  naturally 
expected given the hierarchical 
structure of the charged lepton masses. 
In this case relatively large values of $\theta_{\odot}$ 
and of $|U_{e3}|^2$ are typically predicted, 
$\tan^2\theta_{\odot} \gtap 0.42$,
$|U_{e3}|^2 \gtap 0.02$, while the atmospheric
neutrino mixing angle $\theta_{\rm atm}$ can 
deviate --- depending on the hierarchy of 
the three $\lambda_{ij}$ --
noticeably from $\pi/4$,
$\sin^22\theta_{\rm atm} \gs 0.95$. 
\item $\lambda_{23} \simeq 1$, 
$\lambda_{12, 13}$ are small: $\lambda_{12, 13} \ls 0.35$.\\
The main consequence of this scenario is 
practically maximal atmospheric neutrino mixing 
$\sin^2 2 \theta_{\rm atm} \simeq 1$.
\item All $\lambda_{ij}$ are large: $\lambda_{ij} \gs 0.4$.\\
No interesting correlation of parameters or preferred values of the 
observables is found in this case. 
\end{itemize}
We shall focus in what follows on the 
first two possibilities. 
To estimate the form of the 
relevant matrices, we can  
use the characteristic 
quantity $\lambda \simeq 0.20 - 0.25$, 
which typically appears when the 
magnitude of the angles in $U_L$ 
is constrained. Numerically, we also have 
\be 
m_\mu \sim m_\tau~\lambda^2 \mbox{ and } m_e \sim m_\tau~\lambda^6~,  
\ee
a fact which makes it possible to express the elements of the 
charged lepton mass matrix in terms of powers of $\lambda$. 
Recall that $m_\ell$ is in general diagonalized by a bi--unitary 
transformation defined by $U_L$ and $U_R$. Since only $U_L$ appears in 
the PMNS matrix, we will analyzed two mass matrices associated with the 
charged leptons. The first one is $m_\ell m_\ell^\dagger$, which is 
diagonalized by $U_L$. The second one is defined as 
$U_L m_\ell^{\rm diag} U_L^T$, which would correspond to the charged lepton 
mass matrix if it was symmetrical, a possibility which can occur in certain 
GUT models.


\subsubsection{\label{sec:smla} Small $\lambda_{ij}$}


  The characteristic value of $\lambda_{12}$ 
is approximately 0.23. For the other 
two $\lambda_{ij}$ there 
are basically two possibilities \cite{FPR}. 
Both $\lambda_{23}$ and  $\lambda_{13}$
can be much smaller than $\lambda_{12}$:
$\lambda_{12} = \lambda$, $\lambda_{13} = A~\lambda^3$ and 
$\lambda_{23} = B~\lambda^2$. Here, the parameters $A,B$ are positive 
numbers of order one.

The charged lepton mass matrices $m_\ell m_\ell^\dagger$ and
$m_\ell = m_\ell^T$ have the form:
\be
m_\ell m_\ell^\dagger \sim m_\tau^2 
\left( 
\bad 
\lambda^6 & \lambda^5 & \lambda^3 \\[0.3cm]
\lambda^5 & \lambda^4 & \lambda^2 \\[0.3cm]
\lambda^3 & \lambda^2 & 1 
\ea 
\right)~,~~~
U_L~m_\ell^{\rm diag}~U_L^T \sim m_\tau 
\left( 
\bad 
\lambda^4 & \lambda^3 & \lambda^3 \\[0.3cm]
\lambda^3 & \lambda^2 & \lambda^2 \\[0.3cm]
\lambda^3 & \lambda^2 & 1 
\ea 
\right)~.
\ee 
The expressions for the neutrino mixing 
parameters read \cite{FPR} 
%
\bea \D \label{eq:wolfmix}
\ts \simeq 1 - 2\sqrt{2}~\lambda + 4 \lambda^2 
- 2 \sqrt{2} (2  - A + B) \lambda^3 ~,\\[0.3cm]\D
|U_{e3}| \simeq \left|\frac{\lambda}{\sqrt{2}} + 
\frac{A - B}{\sqrt{2}}\lambda^3 \right|~,\\[0.3cm]\D
\sa  \simeq 1 - \frac{(1  + 4B)^2}{4} \lambda^4~.
\eea
Thus, to leading order, the following correlations 
hold: 
\be \label{eq:wolfcor}
\ts \simeq 1 - 4~|U_{e3}|~(1 - 2|U_{e3}|) -16~ |U_{e3}|^3~,~~
\sa \simeq 1 - (1 + 4B)^2 ~|U_{e3}|^4~.  
\ee
The first relation implies that 
future more precise measurements on the 
magnitude of $|U_{e3}|$ 
can confirm or disfavor, or even rule out,
this scenario (see also \cite{FPR,roma}).\\

  The second possibility corresponds to
$\lambda_{12}$ and $\lambda_{23}$ 
having similar magnitudes, 
say, $\lambda_{12} = \lambda$ and $\lambda_{23} = B~\lambda/2$, but
$\lambda_{13} = A~\lambda^3$. 
One finds in this case 
\be
m_\ell m_\ell^\dagger \sim m_\tau^2 
\left( 
\bad 
\lambda^6 & \lambda^4 & \lambda^3 \\[0.3cm]
\lambda^4 & \lambda^2 & \lambda \\[0.3cm]
\lambda^3 & \lambda & 1 
\ea 
\right)~,~~~
U_L~m_\ell^{\rm diag}~U_L^T \sim m_\tau 
\left( 
\bad 
\lambda^4 & \lambda^3 & \lambda^3 \\[0.3cm]
\lambda^3 & \lambda^2 & \lambda \\[0.3cm]
\lambda^3 & \lambda & 1 
\ea 
\right)
\ee 
%
and
\bea \D \label{eq:mix2} 
\ts \simeq 
1 - 2 \sqrt{2}~\lambda + (4 - \sqrt{2} B) \lambda^2 
- \left(4 \sqrt{2} - 2 \sqrt{2} A - 4 B 
- \frac{B^2}{2 \sqrt{2}} \right) ~\lambda^3 
~,\\[0.3cm]\D
|U_{e3}| \simeq \left|\frac{\lambda}{\sqrt{2}} - 
\frac{B}{2\sqrt{2}}\lambda^2 \right|~,\\[0.3cm]\D
\sa  \simeq 1 - B^2~\lambda^2~.
\eea
%
A similar correlation between \ts{} and $|U_{e3}|$ 
as in the previous case holds
(see Eq.\ (\ref{eq:wolfcor})), but now 
the deviations from 
maximal atmospheric neutrino 
mixing can be larger: 
$\sa \simeq 1 - 2B^2~|U_{e3}|^2$. 


\subsubsection{\label{sec:lam23}$\lambda_{23} \simeq 1$ and 
$\lambda_{12,13}$ small}


   The case of $\lambda_{23} \simeq 1$ and $\lambda_{12,13}$ being 
small is counterintuitive, but nevertheless allowed
by the data \cite{FPR}. There are three 
different possibilities for the hierarchy between
 $\lambda_{12}$ and $\lambda_{13}$:
i) $\lambda_{12} \equiv \lambda $, $\lambda_{13} = A~ \lambda^2$, 
ii) $\lambda_{12} \equiv \lambda^2 $, $\lambda_{13} = A~ \lambda$ and 
iii) $\lambda_{12} \equiv \lambda/2 $, $\lambda_{13} = A~ \lambda/2$,
where $A$ is a real parameter of order one. 
All lead to expressions for, and 
relations between, the neutrino mixing 
parameters similar to those found 
in the ``Wolfenstein--case'' and given in 
Eqs.\ (\ref{eq:wolfmix}) and (\ref{eq:wolfcor}). 
The only particularity can occur in the case  
$\lambda_{12} \equiv \lambda/2 $, $\lambda_{13} = A~ \lambda/2$: 
the two terms in the expression for $|U_{e3}|$ 
can cancel and produce $|U_{e3}| \simeq 0$. 

  We give below the form of the corresponding
charged lepton mass matrices
$m_\ell  m_\ell^\dagger$ and
$m_\ell = m_\ell^T$ in each of these three cases.
For $\lambda_{12} \equiv \lambda $ and $\lambda_{13} = A~ \lambda^2$,
we find 
\be
m_\ell m_\ell^\dagger \sim m_\tau^2 
\left( 
\bad 
\lambda^4 & \lambda^2 & \lambda^5 \\[0.3cm]
\lambda^2 & 1 - \lambda^4 & \lambda^7 \\[0.3cm]
\lambda^5 & \lambda^7 & \lambda^4 
\ea 
\right)~,~~~
U_L~m_\ell^{\rm diag}~U_L^T \sim m_\tau 
\left( 
\bad 
\lambda^4 & \lambda^2 & \lambda^3 \\[0.3cm]
\lambda^2 & 1 - \lambda^4 & \lambda^5 \\[0.3cm]
\lambda^3 & \lambda^5 & \lambda^2 
\ea 
\right)~.
\ee 
%
In the case of $\lambda_{12} \equiv \lambda^2 $ and 
$\lambda_{13} = A~ \lambda$,  we get:
\be
m_\ell m_\ell^\dagger \sim m_\tau^2 
\left( 
\bad 
\lambda^2 & \lambda & \lambda^6 \\[0.3cm]
\lambda & 1 - \lambda^2 & \lambda^7 \\[0.3cm]
\lambda^6 & \lambda^7 & \lambda^4 
\ea 
\right)~,~~~
U_L~m_\ell^{\rm diag}~U_L^T \sim m_\tau 
\left( 
\bad 
\lambda^2 & \lambda & \lambda^4 \\[0.3cm]
\lambda & 1 - \lambda^2 & \lambda^5 \\[0.3cm]
\lambda^4 & \lambda^5 & \lambda^2 
\ea 
\right)~.
\ee 
%
Finally, if 
$\lambda_{12} \equiv \lambda/2 $ and $\lambda_{13} = A~ \lambda/2$, 
one finds 
\be
m_\ell m_\ell^\dagger \sim m_\tau^2 
\left( 
\bad 
\lambda^2 & \lambda & \lambda^5 \\[0.3cm]
\lambda & 1 - \lambda^2 & \lambda^6 \\[0.3cm]
\lambda^5 & \lambda^6 & \lambda^4 
\ea 
\right)~,~~~
U_L~m_\ell^{\rm diag}~U_L^T \sim m_\tau 
\left( 
\bad 
\lambda^2 & \lambda & \lambda^3 \\[0.3cm]
\lambda & 1 - \lambda^2 & \lambda^4 \\[0.3cm]
\lambda^3 & \lambda^4 & \lambda^2 
\ea 
\right)~.
\ee 
%
Note that the largest entry is always the $22$ element.


\subsection{\label{sec:lelmlt} The 
 $L_e - L_\mu - L_\tau$ Flavor Symmetry}


 We shall consider next 
the possibility that 
maximal and/or zero mixing
in $U_{\nu}$ are a consequence of 
the flavor symmetry 
associated with the conservation of 
non--standard lepton charge \cite{lelmlt,lelmlt1}
\[ 
L'= L_e - L_\mu - L_\tau~.
\] 
The most general neutrino Majorana mass matrix 
obeying this symmetry reads \cite{lelmlt} (see also \cite{CNLeungSP83})
\be \label{eq:mnul3l2l1}
m_\nu = m_0 \,  
\left( 
\bad 
0 & -\cos \theta & \sin \theta \\[0.2cm]
\cdot  & 0 & 0 \\[0.2cm]
\cdot & \cdot & 0 
\ea 
\right)~,
\ee
%
where $m_0$ denotes the typical neutrino 
mass scale in this scheme, $m_0 = \sqrt{2\dma}$. 
The fact that atmospheric neutrino mixing 
is observed to be relatively large implies that  
$|\cos \theta| \sim |\sin \theta|$. 
The neutrino mass matrix  
Eq.\ (\ref{eq:mnul3l2l1})  
predicts \cite{lelmlt} 
a neutrino mass spectrum
with inverted hierarchy, 
$m_3 = 0$ (since its rank is 2), 
maximal solar neutrino mixing, $|U_{e3}| = 0$, 
$R = 0$ and atmospheric neutrino mixing 
corresponding to $\theta_{23}^\nu = \theta$. 
Without applying further symmetries, 
e.g., a $\mu - \tau$ symmetry in the 
neutrino sector \cite{mutau}, 
the atmospheric mixing is not maximal, 
and furthermore,
 {\it a priori} is unconstrained. 
Replacing a zero entry with a small 
entry proportional to $\epsilon^2$ 
will generate a non--zero \dms{}, 
according to $R \propto \epsilon^2$. 
We are considering only the 
case of adding one such non--zero entry 
(similar results occur for two or more), 
since this is the simplest possibility. 
In this case one also finds interesting relations 
for the $CP$--violating observables, 
see \cite{FPR} and below. 
Table \ref{tab:break1} summarizes 
the possibilities for adding one entry in $m_{\nu}$, 
which breaks the $L_e - L_\mu - L_\tau~$ symmetry. 
As can be seen from Table \ref{tab:break1}, 
for all cases the corrections to 
$|U_{e3}|= 0$ and to 
$\ts = 1$ are at most proportional 
to $R$ and therefore especially for $\ts$ they are negligible.\\ 
 
\begin{table}[h]
\begin{center}
\begin{tabular}{|c|c|c|c|c|} 
\hline
Matrix $m_{\nu}/m_0$ & $R$ & $m_3/m_0$ & $|U_{e3}|$ & \ts \\ \hline \hline
$ \left( \bad 
 \epsilon^2 & c & s \\
\cdot & 0 & 0 \\ 
\cdot & \cdot & 0 
\ea \right)$ 
& $2~\epsilon^2$ & 0 & 0 & $1 + \frac{\D R}{\D 2} $ \\ \hline
$ \left( \bad 
0 & c & s \\
\cdot &  \epsilon^2 & 0 \\ 
\cdot & \cdot & 0 
\ea \right)$ 
& $2~c^2~\epsilon^2$ & 
$\frac{\D R}{\D 2} \, 
\frac{\D \tan \theta_{\rm atm}}{\D \cos \theta_{\rm atm}}$ 
 &  $\frac{\D R}{\D 2} \, \tan \theta_{\rm atm}$ 
& $1 - \frac{\D R}{\D 2} $ \\ \hline
$ \left( \bad 
0 & c & s \\
\cdot & 0 & 0 \\ 
\cdot & \cdot &  \epsilon^2 
\ea \right)$ & 
\multicolumn{4}{c|}{same as above with $c \leftrightarrow s$} \\ \hline
$ \left( \bad 
0 & c & s \\
\cdot & 0 &  \epsilon^2  \\ 
\cdot & \cdot & 0
\ea \right)$ 
& $4~c~s~\epsilon^2$  & $\frac{\D R}{\D 2}$ &  
$\frac{\D R}{\D 2} \, \cot \theta_{\rm atm}$ 
& $1 - \frac{\D R}{\D 2} $ \\ \hline 
\end{tabular}
\caption{\label{tab:break1} Possible ways of breaking 
of the flavor symmetry 
implying the conservation of $L_e - L_\mu - L_\tau$,
by adding in $m_{\nu}$ one small (but non-zero) entry $m_0 \epsilon^2$ 
which generates non--vanishing 
$R \equiv \dms/\dma \sim 0.04$. 
Shown are the expressions for the ratio $R$, the 
smallest neutrino mass $m_3$ 
(in units of $m^2_0 = 2|\dma|$), 
and the mixing parameters $|U_{e3}|$ and \ts. The 
``large'' $e\mu$ and $e\tau$ entries of $m_{\nu}/m_0$ 
are denoted by $c = \cos \theta$ and $s = \sin \theta$. 
In all cases except the first, 
one has $\tan^2 \theta_{23}^\nu =  \tan^2 \theta 
+ {\cal O}(\epsilon^4)$; in the first case the relation  
$\tan^2 \theta_{23}^\nu = \tan^2 \theta$ holds exactly.}
\end{center}
\end{table}

   Let us stress again that in the general case of 
$L_e - L_\mu - L_\tau$ conservation,
the atmospheric neutrino mixing is unconstrained, i.e., $U_\nu$ 
has the form 
\be
U_\nu = 
\left(
\bad
\frac{\D 1}{\D \sqrt{2}} & \frac{\D 1}{\D \sqrt{2}} & 0 \\[0.3cm]
-\frac{\D \cos \theta^\nu_{23}}{\D \sqrt{2}} & 
 \frac{\D \cos \theta^\nu_{23}}{\D \sqrt{2}} &  \sin \theta_{23}^\nu\\[0.3cm]
  \frac{\D \sin \theta^\nu_{23}}{\D \sqrt{2}} & 
 -\frac{\D \sin \theta^\nu_{23}}{\D \sqrt{2} }
& \cos \theta_{23}^\nu
\ea
\right)~. 
\ee
It holds  
$\tan \theta_{23}^\nu = (m_{\nu})_{e\tau}/(m_{\nu})_{e\mu}$ \cite{lelmlt} and 
with no additional contribution 
from the charged lepton sector 
this angle would correspond to 
the atmospheric neutrino mixing angle. 
In order to find the allowed 
forms of the charged lepton mixing 
matrix in this case, we should therefore redo 
the analysis performed in \cite{FPR}
with $\theta_{23}^\nu$
not necessarily equal to $\pi/4$. 
This will be done in the next Section. 


\section{\label{sec:gen} Generalization to $\theta_{23}^\nu \neq \pi/4$}


   For illustrative reasons we will  
consider first the case of $CP$ invariance
and analyze after that the 
more interesting and realistic case 
of $CP$ nonconservation. 


\subsection{\label{sec:CPC} The Case of $CP$ Invariance}

\begin{figure}[p]
\begin{center}
\vspace{-3cm}
\epsfig{file=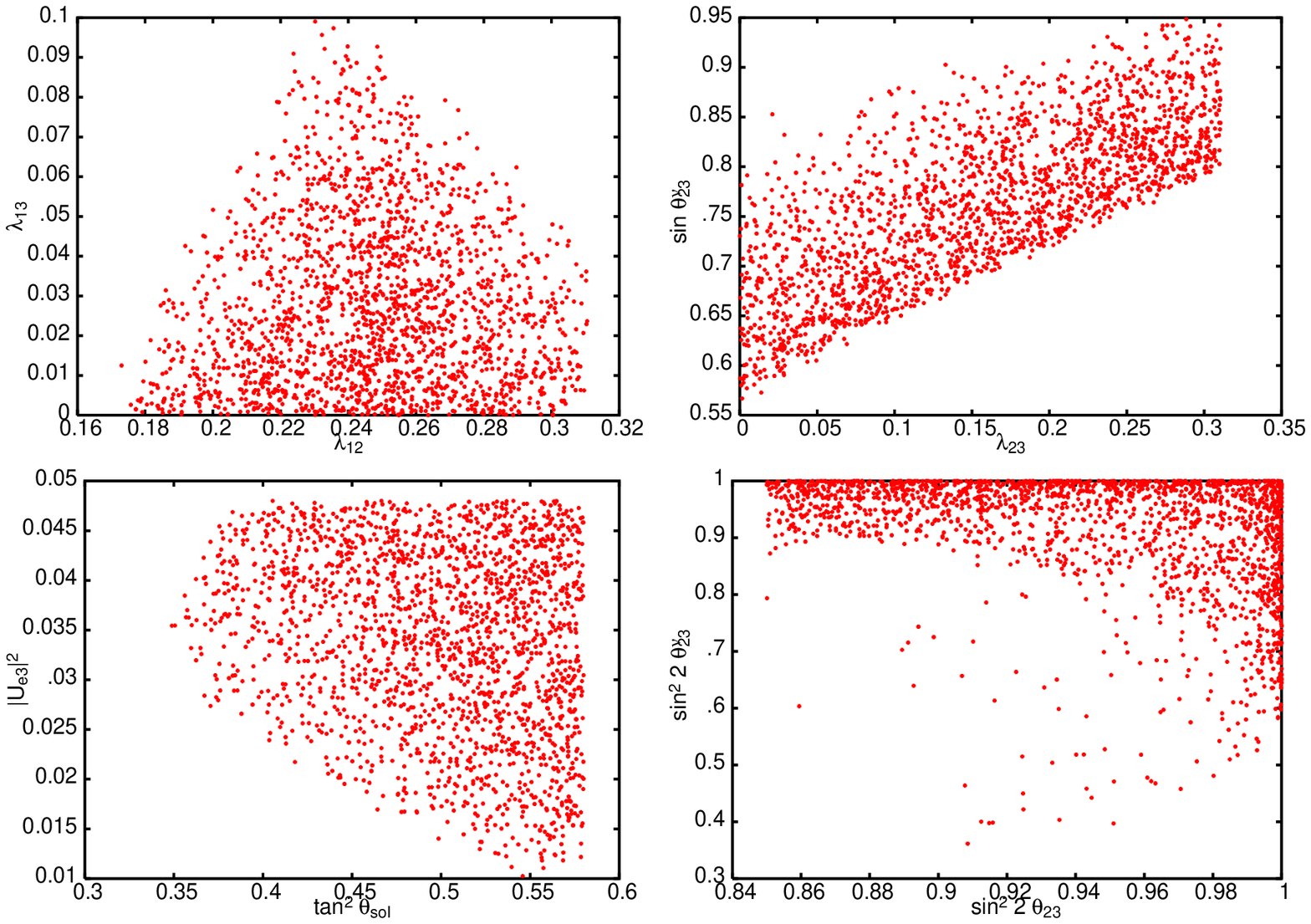,width=19cm,height=24cm}
\vspace{-10cm}
\caption{\label{fig:leptoCPCsmla}
Scatter plot of some of the $\lambda$ parameters, $\sin \theta_{23}^\nu$ 
and the neutrino mixing observables 
for the 3$\sigma$ allowed ranges of values of the 
neutrino mixing parameters given 
in Eq.\ (\ref{eq:range}). All $\lambda_{ij}$ are small and 
conservation of $CP$ is assumed.}
\end{center}
\end{figure}

  We repeated the analysis performed in \cite{FPR}, 
the only difference being that 
$\tan \theta_{23}^\nu$ was
allowed to vary in a certain range. 
Since $\tan \theta_{23}^\nu$ is the ratio of the 
two ``large'' entries in $m_{\nu}$
expected to be $\sim m_0$,
$\tan \theta_{23}^\nu$ was allowed 
to take values between $1/3$ and $3$.  
As in the case of $\theta_{23}^\nu = \pi/4$ analyzed in \cite{FPR}, 
the parameter regions favored by the data were 
those corresponding to
all $\lambda_{ij}$ being ``small'', all $\lambda_{ij}$ being ``large'' 
and $\lambda_{23} \simeq 1$ with the other two $\lambda_{ij}$ ``small''. 
The same parameter regions are present 
also if $CP$ is not conserved, as is discussed below.
Only the case of all $\lambda_{ij}$ being small 
exhibits interesting correlations 
of neutrino mixing observables and  
we display some of them in Fig.\ \ref{fig:leptoCPCsmla}. 

We constrained $\lambda_{ij}$ to vary in the interval 
 $ 0 \leq \lambda_{ij} \leq 1/\sqrt{10}$, 
and required that the neutrino 
mixing observables 
are inside their 3$\sigma$ allowed 
ranges given in Eq.\ (\ref{eq:range}). 
Again, $\lambda_{12} \equiv \lambda$ is centered around $0.24$.  
For $\lambda_{23}$ there are the possibilities 
of $\lambda_{23} = {\cal O}(\lambda)$, in which case typically 
$\theta_{23}^\nu \gs \pi/4$,
and of $\lambda_{23} = {\cal O}(\lambda^2)$, leading 
typically to $\theta_{23}^\nu \ls \pi/4$.

   For $\theta_{23}$ we encountered in Ref.\ \cite{FPR} 
the typical correlation $\sin^2 2 \theta_{\rm atm} 
\sim 1 - 2~|U_{e3}|^2$ or $\sin^2 2 \theta_{\rm atm} 
\sim 1 - |U_{e3}|^4$, which implies that at most 
$\sin^2 2 \theta_{\rm atm} \gs 0.9$ \cite{FPR}. As seen in Fig.\ 
\ref{fig:leptoCPCsmla}, atmospheric neutrino mixing can now 
be anything between its lowest allowed value of 0.85 and 1. 
The parameter $|U_{e3}|^2$ is bounded to be larger than 
approximately  0.01, a limit weaker by a factor 
of two with respect to the case 
of $\theta_{23}^\nu = \pi/4$.

As a typical example, let us choose $\lambda_{12} \equiv \lambda$, 
$\lambda_{23} = B \, \lambda^2$ and small 
$\lambda_{13} = A \, \lambda^3$. We find 
\bea \label{eq:mixsmla1}
\ts \simeq 1 - 4 \, \cos \theta_{23}^\nu \, \lambda + 
8\,\cos^2 \theta_{23}^\nu \, \lambda^2~, \\[0.3cm]
|U_{e3}| \simeq |\sin \theta_{23}^\nu \, \lambda +  
(A - B) \cos \theta_{23}^\nu \, \lambda^3|~, \\[0.3cm]
\tan^2 \theta_{\rm atm} \simeq \tan^2  \theta_{23}^\nu 
- \frac{\D \sin  \theta_{23}^\nu}{\D \cos ^3 \theta_{23}^\nu}~
(4 B + \sin 2 \theta_{23}^\nu) \, \lambda^2,~~\mbox{ or } \\[0.3cm]
 \sa \simeq \sin^2 2 \theta_{23}^\nu - 
\frac{1}{2}(4 B +  \sin 2 \theta_{23}^\nu)~\sin 4 \theta_{23}^\nu \, 
\, \lambda^2 
\eea 
plus higher order terms. Setting $\theta_{12}^\nu$ to $\pi/4$ reproduces 
Eq.\ (\ref{eq:wolfmix}).  
To lowest order in $\lambda$ we have 
$\theta_{\rm atm} =  \theta_{23}^\nu$, i.e.,
the atmospheric neutrino mixing angle corresponds to the 
23 mixing angle in $U_{\nu}$. 
We explicitly gave also the expression for $\tan^2 \theta_{\rm atm}$ 
to address the question of whether 
$\theta_{\rm atm} > \pi/4$ or $\theta_{\rm atm} < \pi/4$. 
This ambiguity is part of the 
``eightfold degeneracy'' from which the interpretation of the results 
of future long baseline neutrino experiments may suffer \cite{8fold}.\\

There would be physically interesting consequences for 
the oscillations of atmospheric neutrinos,
if $\theta_{\rm atm}$ differs significantly from $\pi/4$.
A value of $\theta_{\rm atm} \gs \pi/4$ would imply 
larger (than in the case $\theta_{\rm atm} < \pi/4$)
probabilities of the subdominant 
$\nu_e \rightarrow \nu_{\mu}$ ($\bar{\nu}_e \rightarrow \bar{\nu}_{\mu}$) and  
$\nu_\mu \rightarrow \nu_{e}$ ($\bar{\nu}_\mu \rightarrow \bar{\nu}_{e}$)
oscillations of the multi--GeV atmospheric neutrinos 
$\nu_e$ ($\bar{\nu}_e$) and $\nu_\mu$ ($\bar{\nu}_\mu$) \cite{ChMP98}.
For sufficiently large values of $\theta_{13}$, 
$\sin^22\theta_{13} \gtap 0.05$,
these oscillations can be strongly enhanced by the Earth matter 
and their effects on Nadir angle distributions of the 
multi--GeV samples of atmospheric neutrino 
$e$--like and $\mu$--like 
(or $\mu^-$ and $\mu^+$) samples of events 
could be observable in the operating and 
future atmospheric neutrino detectors \cite{JBSP203}. 
For non--maximal atmospheric neutrino mixing $\theta_{\rm atm} < \pi/4$,
the  subdominant $\nu_e \rightarrow \nu_{\mu}$ 
and  $\nu_\mu \rightarrow \nu_{e}$
oscillations driven by $\dms$ 
can lead to relatively large effects in the sub--GeV sample of 
atmospheric $e$--like events measured in the SuperKamiokande 
experiment \cite{atmt23}.\\
 
 It follows from Eq.\ (\ref{eq:mixsmla1}) that for 
$\theta_{23}^\nu = \pi/4$, the multiplication 
of $U_\nu$ with $U_{\rm lep}^\dagger$ drags the atmospheric neutrino 
mixing towards $\theta_{\rm atm} < \pi/4$. 
The presence of $CP$--violating phases can 
change this behavior, as will be shown below.

 The quadratic correction to $\sa = \sin^2 2 \theta_{23}^\nu$ 
vanishes for $\theta_{23}^\nu = \pi/4$, 
the next term being quartic in $\lambda$. 
Since  to lowest order in $\lambda$ 
we have $\sin \theta_{23}^\nu \simeq \sin \theta_{\rm atm}$, we can write
using Eq.\ (\ref{eq:mixsmla1}) 
\be \label{eq:cornew}
\ts \simeq 1 - 4 \, |U_{e3}| \, \cot \theta_{\rm atm}~.
\ee
In \cite{roma} it was 
stated correctly that the lowest order relation 
$\ts \simeq 1 - 4~|U_{e3}|$ 
can make this scenario problematic
when improved limits on $|U_{e3}|$ will be available. 
Allowing $\theta_{23}^\nu$ 
to differ from $\pi/4$ 
makes it easier to satisfy 
the conflicting requirements 
in the scheme discussed 
of small $|U_{e3}|$ and $\ts \simeq 0.40$. 
Note, though, that the constraints on the 
atmospheric neutrino mixing still 
have to be satisfied. 

To sum up, the fact that 
$\theta_{23}^\nu$ has to be 
close to $\pi/4$, results in 
very similar correlations between 
the neutrino mixing observables, that is 
$\ts \sim 1 - 4~ |U_{e3}|$ and 
$\sa \sim 1 - |U_{e3}|^2$ ($\sa \sim 1 - |U_{e3}|^4$) for 
$\lambda_{23} \sim \lambda$ ($\lambda_{23} \sim \lambda^2$). 
Nevertheless, better limits on $|U_{e3}|$ will not automatically 
lead to a conflict with  $\ts \simeq 0.40$, 
unless the atmospheric neutrino mixing turns out to be very 
close to maximal. For instance, for the minimally allowed 
at 1$\sigma$ value of $\sa = 0.95 \simeq \sin 2 \theta_{23}^\nu$, 
we have $\ts \simeq 1 - 5~|U_{e3}|$, to be compared with 
the lowest order relation in the case of maximal $\theta_{23}^\nu$: 
$\ts \simeq 1 - 4~|U_{e3}|$. 
Thus, the presence of 
a non--maximal $\theta_{23}^\nu$ allows for smaller values 
of $|U_{e3}|$ while at the same time being in agreement with the observed 
non--maximal solar neutrino mixing. This is basically independent on the 
precise form of $U_L$, as long as the angles in $U_L$ are all small, and holds 
also in the case of $CP$ nonconservation.

\subsection{\label{sec:CPV}Violation of $CP$ }

\subsubsection{\label{sec:CPVgen}General Considerations}

As was shown in \cite{FPR},
there are six $CP$--violating phases, in general,
from the neutrino and charged lepton sector 
contributing to the PMNS matrix 
$\pmns = U_L^\dagger U_\nu$. 
Using the general description of 
unitary matrices given in \cite{PPR}, one can show that 
\be \label{eq:pmnscpv}
\pmns = \tilde{U}_{\rm lep}^\dagger \, P_\nu \, \tilde{U}_\nu \, Q_\nu ~.
\ee
Here, $P_\nu \equiv {\rm diag} (1,e^{i \phi},e^{i \omega})$
and $Q_\nu \equiv {\rm diag} (1,e^{i \rho},e^{i \sigma}) $  
are diagonal phase matrices having 
two phases each, and $\tilde{U}_{\nu, \rm lep}$ are unitary 
matrices, each containing one phase and three angles. Only 
 $\tilde{U}_{\rm lep}$ stems from the diagonalization of the charged 
lepton matrix, all the other three matrices have their origin in the 
neutrino sector. Note that $Q_\nu$ is ``Majorana--like'', i.e., its two 
phases will not affect
the flavor neutrino oscillations. They can enter into 
the expressions of observables associated with 
processes in which the total lepton charge is not conserved,
such as neutrinoless double beta decay, see, e.g., \cite{0vbbCP,BPP1}.  
Finally, the presence of one zero angle in $\tilde{U}_{\nu}$ 
means that the Dirac--like phase 
in $\tilde{U}_{\nu}$ can be set to zero. 
We note in passing that 
if the bimaximal mixing would 
stem entirely from the charged lepton 
sector, the Dirac--like phase in $\tilde{U}_{\rm lep}$ 
would not be physical and all leptonic $CP$ violation 
would stem from the neutrino sector.

  In the case of 3--neutrino mixing under discussion
there are, in principle, three $CP$ violation rephasing invariants.
The first is the standard Dirac one $J_{CP}$ \cite{CJ85}, associated
with the Dirac phase $\delta$ and measured in 
neutrino oscillation experiments \cite{PKSP88}:  
\be 
\label{eq:JCP}
J_{CP} = 
{\rm Im} \left\{ U_{e1} \, U_{\mu 2} \, U_{e 2}^\ast \, U_{\mu 1}^\ast 
\right\}~. 
\ee
%
There are two additional 
invariants, $S_1$ and $S_2$, whose existence is
related to the two Majorana $CP$ violation phases in 
$\pmns$, i.e., to the Majorana nature of massive neutrinos. 
As can be shown, the effective Majorana mass measured in 
$\betabeta$--decay experiments depends, in general, on 
these two invariants 
and not on $J_{CP}$ \cite{BPP1}. 
The invariants $S_1$ and $S_2$ can be chosen as
\footnote{We assume that the fields
of massive Majorana neutrinos satisfy
Majorana conditions which 
do not contain phase factors.} \cite{JMaj87,ASBranco00,BPP1}:
\be
S_1 = {\rm Im}\left\{ U_{e1} \, U_{e3}^\ast \right\}~,~~~~
S_2 = {\rm Im}\left\{ U_{e2} \, U_{e3}^\ast \right\}~.
\label{eq:SCP}
\ee
%
If $U_{e3} = 0$,
the Majorana phases $\alpha$ and $\beta$ 
in $\pmns$ can still induce 
$CP$--violating effects as long as  
${\rm Im}\left\{ U_{e1} \, U_{e2}^\ast \right\} \neq 0$
and ${\rm Im}\left\{ U_{\mu 2} \, U_{\mu 3}^\ast \right\}\neq 0$ 
\cite{ASBranco00}. 

    Turning to the $L_e - L_\mu - L_\tau$ symmetry, we can now 
include the possibility of $CP$ violation \cite{FPR}. 
Suppose, for example, that all non--zero entries in 
$m_\nu$ are complex and consider, e.g., 
the second case in Table \ref{tab:break1}, 
i.e., a small perturbation in the $\mu\mu$ 
entry. The neutrino mass matrix can then be written as 
\bea \label{eq:mnu2}
m_\nu = m_0 
\left( 
\bad 
0 & c \, e^{i \beta} & s \, e^{i \gamma} \\[0.2cm]
\cdot & \epsilon^2 \, e^{i \alpha} & 0 \\[0.2cm]
\cdot & \cdot & 0
\ea
\right) \\[0.3cm] 
= m_0 \, {\rm diag}( e^{i(\beta - \alpha/2)}, e^{i \alpha/2}, 
e^{i(\alpha/2 + \gamma - \beta )}) \, 
\left( 
\bad 
0 & c  & s  \\[0.2cm]
\cdot & \epsilon^2 & 0 \\[0.2cm]
\cdot & \cdot & 0
\ea
\right) \, {\rm diag}( e^{i(\beta - \alpha/2)}, e^{i \alpha/2}, 
e^{i(\alpha/2 + \gamma - \beta )}) ~, 
\eea
%
and the real (inner) matrix can be diagonalized by a real 
orthogonal matrix $O$. 
Following the procedure outlined in Section 4.6 of Ref.\ \cite{FPR}, 
we can identify the phases $\phi$ and $\omega$ present in $P_\nu$, 
Eq.\ (\ref{eq:pmnscpv}), with $(\alpha - \beta)$
and $(\alpha - 2\beta + \gamma)$.
The phase matrix $Q_\nu$ 
is equal to the unit matrix \cite{FPR}. 
  

\subsubsection{\label{sec:CPVfirst}First example}

\begin{figure}[p]
\begin{center}
\epsfig{file=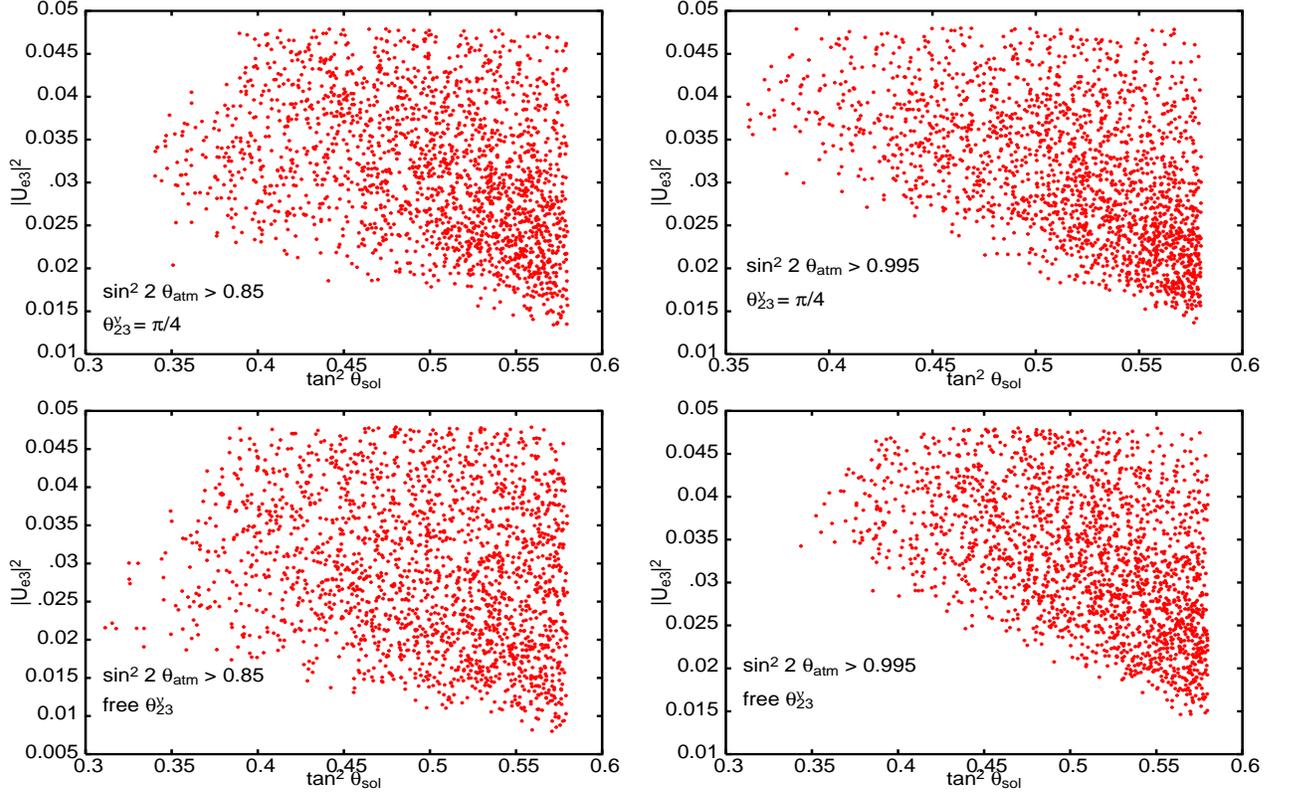,width=19cm,height=24cm}
\vspace{-10cm}
\caption{\label{fig:tsolue3comp}
Scatter plot of \ts{} against $|U_{e3}|^2$ for two different 
limits on \sa{} and for the cases $\theta_{23}^\nu = \pi/4$ and 
free $\theta_{23}^\nu$. A ``CKM--like'' hierarchy of 
the $\lambda_{ij}$ is assumed and 
it is required that the neutrino mixing 
parameters $\ts$ and $|U_{e3}|$
have values in their respective
3$\sigma$ allowed 
intervals, given in Eq.\ (\ref{eq:range}).}
\end{center}
\end{figure}

  Let us express now the neutrino mixing observables 
in terms of the parameters in $U_\nu$ and $U_L$.
As in the case of $CP$ conservation
we choose $\lambda_{12} \equiv \lambda$, 
$\lambda_{23} =  B \, \lambda^2$ and 
$\lambda_{13} = A \, \lambda^3$. 
For the mixing parameters we get
\bea \label{eq:cor1mix}
\ts \simeq 1 - 4\, c_\phi \, \cos \theta_{23}^\nu \, \lambda + 
8 \, c_\phi^2 \, \cos^2 \theta_{23}^\nu \, \lambda^2 ~,\\[0.3cm]
|U_{e3}| \simeq |\sin \theta_{23}^\nu \, \lambda - 
(B \, c_{\omega - \phi} - A \, c_{\omega - \phi - \psi}) 
\, \lambda^3|~,\\[0.3cm] 
\tan^2 \theta_{\rm atm} \simeq \tan^2  \theta_{23}^\nu 
- \frac{\D \sin  \theta_{23}^\nu}{\D \cos ^3 \theta_{23}^\nu}~
(4 B \, c_{\omega - \phi}+ \sin 2 \theta_{23}^\nu) \, \lambda^2~
,~~\mbox{ or } \\[0.3cm]
\sa \simeq \sin^2 2 \theta_{23}^\nu - 
\frac{1}{2}(4 B \, c_{\omega - \phi} 
+  \sin 2 \theta_{23}^\nu)~\sin 4 \theta_{23}^\nu \, 
\, \lambda^2 ~.
\eea
%
Here $\psi$ is the Dirac--like phase in $\tilde{U}_{\rm lep}$ 
and we used the obvious notation $c_\phi = \cos \phi$, etc.  
Comments similar to those in the $CP$--conserving case discussed above 
can be made. In particular, we have now a similar correlation 
between solar neutrino mixing and $|U_{e3}|$:
\be \label{eq:corCPV}
\ts \simeq 1 - 4 \, \cos \phi \, \cot \theta_{\rm atm} \, |U_{e3}| ~.
\ee   
If $|U_{e3}|$ is relatively small,
say $|U_{e3}| \ltap 0.10$,
the phase $\phi$ should 
be close to $0$ or $2\pi$ in order to  
be compatible with the data 
on $\ts$.  In the previous Section we mentioned the possibility that  
non--maximal $\theta_{23}^\nu$ might resolve the 
possible conflict between a relatively 
small value of $|U_{e3}|$ and 
significantly non--maximal solar neutrino mixing. 
Such a possibility holds when atmospheric 
neutrino mixing is not too close to maximal. This is illustrated in 
Fig.\ \ref{fig:tsolue3comp}, where 
the correlation between \ts{} and $|U_{e3}|^2$
is shown in the case of $CP$ nonconservation for two different 
limits on \sa{} and for the cases of $\theta_{23}^\nu = \pi/4$ and 
of ``free'' $\theta_{23}^\nu$, $1/3 \leq \tan \theta_{23}^\nu \leq 3$. 
We limited $\lambda_{12} \equiv \lambda \le 1/\sqrt{10}$ 
and chose $\lambda_{23} = B~\lambda^2$ and 
$\lambda_{13} = A~\lambda^3$, with $A,B$  varying 
between $1/\sqrt{3}$ and $\sqrt{3}$. 
It is seen that in the case of ``free'' $\theta_{23}^\nu$ 
and the current limit 
of $\sa > 0.85$, smaller values of both \ts{} and $|U_{e3}|$ are 
possible, whereas 
in the case of $\sa > 0.995$ no difference 
exists between the two cases. 
In fact, the lower limit on $|U_{e3}|^2$ 
is roughly a factor of two weaker when $\theta_{23}^\nu$ is not 
$\pi/4$ ($|U_{e3}|^2 \gs 0.013$ and $\gs 0.007$, respectively). 
Note that the exact expression for the observables
were used to produce the plots
in Fig.\ \ref{fig:tsolue3comp}
(and not the leading order relations, 
Eq.\ (\ref{eq:corCPV})) .\\ 

 It proves useful to derive an expression for the 
difference between $S_1$ and $S_2$ in the case of 
$\rho = \sigma = 0$. As shown in \cite{FPR} 
(see also the preceding discussion), 
these two phases vanish if, e.g., the 
neutrino mass matrix 
has the form given in Eq.\ (\ref{eq:mnu2}). 
When all $\lambda_{ij}$ are 
small, $S_1 - S_2$ will take its maximal value 
for $\lambda_{12} \equiv \lambda$, $\lambda_{23} = B \, \lambda$ and 
$\lambda_{13} =A  \, \lambda$, i.e., when all three $\lambda_{ij}$
have similar magnitudes. Then one finds:
\be
\left. (S_1 - S_2) \right|_{\rho = \sigma = 0} \simeq 
\sqrt{2}A~s_{\omega - \phi - \psi}~\lambda^2 
+ \sqrt{2}B~(A^2 - 1)~s_{\omega - \phi}~\lambda^3~.
\ee
%
A definite hierarchy between the different $\lambda_{ij}$ will be reflected 
in this relation by, e.g., replacing $B$ with $B~\lambda$ for 
$\lambda_{23} = {\cal O}(\lambda^2)$ and so on. Note that for 
the case we are interested in, $\rho = \sigma = 0$, the two invariants 
$S_1$ and $S_2$ are equal at least to order $\lambda$. 

   We will give next 
expressions for the $CP$--violating rephasing invariants.
For $J_{CP}$ we find 
\be \label{eq:cor1JCP}
J_{CP} \simeq \frac{1}{2} \cos \theta_{23}^\nu \, \sin^2 \theta_{23}^\nu \, 
s_{\phi} \, \lambda + {\cal O}(\lambda^3)~. 
\ee
%
We therefore identify (to leading order in $\lambda$) 
$\phi$ as the Dirac phase, measurable in neutrino oscillation 
experiments. 

  As an example for the correlations between 
$CP$--conserving and $CP$--violating 
observables, consider the following case: 
if $|U_{e3}|$ is close to the minimal value in
its allowed range, 
$\cos\phi$ has to be close to one, and thus $\phi$ close to zero or $2\pi$,  
in order for 
the relation $\ts \simeq 1 - 4\cos\phi~|U_{e3}|$
to be compatible with measured value of $\ts$.
This implies that $J_{CP} \propto \sin \phi$ will be 
additionally suppressed. 

Recall that the charged lepton sector 
contributes with only one phase to the PMNS matrix. 
The hierarchical structure of $U_L$, which is under discussion here, 
will suppress the effects of this phase and the main dependence of the 
$CP$--violating observables can be expected to come from 
the neutrino sector. To check this consideration, suppose 
now that the neutrino mass matrix $m_{\nu}$ conserves $CP$. 
In our framework this would mean $\phi = 0$.  
Since to leading order in $\lambda$, $J_{CP} \propto \lambda \, \sin \phi$, 
the $CP$--violating effects in neutrino oscillations 
will depend on the phase $\psi$ in $\tilde{U}_{\rm lep}$ 
and appear only at the next higher order, 
which turns out to be $\lambda^3$.\\ 

  The rephasing invariants associated with the 
Majorana nature of massive neutrinos 
are given by: 
\be \label{eq:S1}
S_1 \simeq \frac{1}{\sqrt{2}} \, \sin \theta_{23}^\nu \, 
s_{\phi + \sigma} \, \lambda + \frac{1}{\sqrt{2}} 
\, \sin \theta_{23}^\nu \, \cos \theta_{23}^\nu \, s_\sigma \, \lambda^2 
\ee
and 
\be \label{eq:S2}
S_2 \simeq \frac{1}{\sqrt{2}} \, \sin \theta_{23}^\nu \, 
s_{\phi -\rho + \sigma} \, \lambda + \frac{1}{\sqrt{2}} 
\, \sin \theta_{23}^\nu \, \cos \theta_{23}^\nu \, s_{\rho - \sigma} 
\, \lambda^2 ~.
\ee
%
In the case of $\rho = \sigma = 0$, 
which is realized if, e.g., the 
neutrino mass matrix has the form of Eq.\ (\ref{eq:mnul3l2l1}) modified by
one additional small term (as summarized in Table \ref{tab:break1}),  
to a good precision 
(i.e., up to corrections $\sim \lambda^3$)
we have
\be \label{eq:S1S2JCP}
S_1 \simeq S_2 \simeq  \frac{2\sqrt{2}}{\sin 2\theta_{23}^\nu} \, J_{CP} 
\simeq \frac{2\sqrt{2}}{\sin 2\theta_{\rm atm}} \, J_{CP} ~,
\ee
where we have used the fact that $\theta_{\rm atm} \simeq \theta_{23}^\nu$.\\

  We will discuss next
whether in the scheme we are analyzing one has 
$\theta_{\rm atm} > \pi/4$ or  $\theta_{\rm atm} < \pi/4$. 
It is instructive to consider the 
difference of $|U_{\mu 3}|$ and $|U_{\tau 3}|$, which reads 
\bea 
|U_{\mu 3}| - |U_{\tau 3}| \simeq 
\sin \theta_{23}^\nu - \cos \theta_{23}^\nu - \frac{\D \lambda^2}{\D 2}~
\left( 
\sin \theta_{23}^\nu + 2B \, c_{\omega - \phi}~
(\cos \theta_{23}^\nu + \sin \theta_{23}^\nu)
\right) \\[0.3cm]
\stackrel{\theta_{23}^\nu \ra \pi/4}{\D \ra} 
- \frac{\D \lambda^2}{\D 2\sqrt{2}}~
\left( 
1 + 4 B \, c_{\omega - \phi}
\right) ~.
\eea
%
Thus, for $\theta_{23}^\nu = \pi/4$ and in the presence of 
$CP$ violation, it is possible  
to have both cases, 
$\theta_{\rm atm} > \pi/4$ and $\theta_{\rm atm} < \pi/4$. 
For $\theta_{23}^\nu = \pi/4$ and $CP$ conservation we have 
$|U_{\mu 3}| < |U_{\tau 3}|$ and thus typically $\theta_{\rm atm} < \pi/4$. 
 We will elaborate further on this 
issue in the next 
Subsection. 

 We will comment next on the effective Majorana mass \meff{} measurable  
in \onbb. This process is sensitive to the absolute value of the 
$ee$ element of the neutrino mass matrix in the basis 
in which the charged lepton mass matrix is real and diagonal. In general, 
$\meff$ is independent of the atmospheric neutrino mixing angle 
$\theta_{23}$ and on the Dirac phase $\delta$ \cite{BPP1}. 
The relation between the invariants corresponding to 
the Dirac and Majorana phases, as encoded in Eq.\ (\ref{eq:S1S2JCP}), 
implies, however, that the expression for $\meff$ will depend on 
$J_{CP}$ which determines 
the magnitude of $CP$ violation effects
in neutrino oscillations \cite{PKSP88}. 
In the case of inverted hierarchy of neutrino masses, predicted by 
the $L_e - L_\mu - L_\tau$ symmetry, 
the effective Majorana mass can be 
written as \cite{lelmlt} (see also, e.g., 
\cite{BiPet87,0vbbCP,BPP1}) 
\be 
\meff \simeq \sqrt{\dma}~
\left| U_{e1}^2 - U_{e2}^2 
\right| ~.
\ee
%
In the case under discussion 
one finds $U_{e1}^2 - U_{e2}^2 
\simeq 2 \cos \theta_{23}^\nu \, e^{i \phi} \, 
\lambda + {\cal{O}} (\lambda^3)$.  
Using Eqs.\ (\ref{eq:cor1mix}) and (\ref{eq:cor1JCP}), 
one can therefore  express $\meff$ as 
\be \label{eq:meff1}
\meff \simeq \sqrt{\dma}~
\left| 
\cos 2 \theta_\odot + 4i~J_{CP}/\sin^2 \theta_{\rm atm} 
\right|~,
\ee
which contains both, 
the Dirac phase, i.e., $J_{CP}$, and the atmospheric neutrino mixing angle. 
In the case of $\theta_{23}^\nu = \pi/4$ we recover the
expression for $\meff$ found  in \cite{FPR}. Note that this relation between 
the effective mass, $J_{CP}$ and the atmospheric neutrino mixing 
does not depend on where the small perturbation in $m_\nu$ is introduced.  


\subsubsection{\label{sec:sec}Second example}


  If we choose now a weaker hierarchy between
the three $\lambda_{ij}$,
i.e.,  $\lambda_{12} \equiv \lambda$, 
$\lambda_{23} = B \, \lambda$ and 
$\lambda_{13} = A \, \lambda^3$, we find: 
\bea
\ts \simeq 1 - 4\, c_\phi \, \cos \theta_{23}^\nu \, \lambda + 
4 \, ( 2 \, c_\phi^2 \, \cos^2 \theta_{23}^\nu 
- B \, c_\omega \, \sin \theta_{23}^\nu) \, \lambda^2 ~,\\[0.3cm]
|U_{e3}| \simeq |\sin \theta_{23}^\nu \, \lambda - 
B \, c_{\omega - \phi}  \, \cos \theta_{23}^\nu \, \lambda^2|~,\\[0.3cm] 
\tan^2 \theta_{\rm atm} \simeq \tan^2  \theta_{23}^\nu 
- \frac{\D \sin  \theta_{23}^\nu}{\D \cos ^3 \theta_{23}^\nu}~
2 B \, c_{\omega - \phi} \, \lambda \\[0.3cm]
\mbox{ or }
\sa \simeq \sin^2 2 \theta_{23}^\nu - 
2 B \, c_{\omega - \phi} \, \sin 4 \theta_{23}^\nu 
\, \lambda ~,
\eea
%
where we omitted the lengthy expression for the sizable quadratic terms in 
$\tan^2 \theta_{\rm atm}$ and \sa. 
The presence of the linear term in $\lambda$, which 
is proportional to $\sin  \theta_{23}^\nu/\cos ^3 \theta_{23}^\nu$,  
explains why for the choice 
of $\lambda_{23} \propto \lambda$
the angle $\theta_{23}^\nu $ 
has a tendency to be smaller than
$\pi/4$. We mentioned this trend while discussing  
Fig.\ \ref{fig:leptoCPCsmla} and can understand 
this now by noting that 
$\sin  \theta_{23}^\nu/\cos ^3 \theta_{23}^\nu$ decreases when  
$\theta_{23}^\nu $ decreases starting from $\pi/4$. 

 Thus,  when $\theta_{23}^{\nu} > \pi/4$,
it will be possible 
to have $\theta_{\rm atm} < \pi/4$ for $\lambda_{23} \propto \lambda$. 
If $\lambda_{23} \propto \lambda^2$, 
$\theta_{\rm atm}$ prefers to stay above $\pi/4$. 

   In the case of $CP$ violation, 
the same expressions for 
$S_{1,2}$ and $J_{CP}$ as those given 
in the previous Subsection will hold, 
when one formally replaces 
$B$ with $B/\lambda$. As can be seen from 
Eqs.\ (\ref{eq:cor1JCP}) - (\ref{eq:S2}),
to leading order the results are the same. 
In particular, the 
relation Eq.\ (\ref{eq:S1S2JCP}) between 
the three $CP$--violating rephasing invariants
$J_{CP}$, $S_1$ and $S_2$ holds again. 
Moreover, the effective Majorana mass \meff{} 
is given by the same expression (\ref{eq:meff1}). 

\begin{figure}[p]
\begin{center}
\epsfig{file=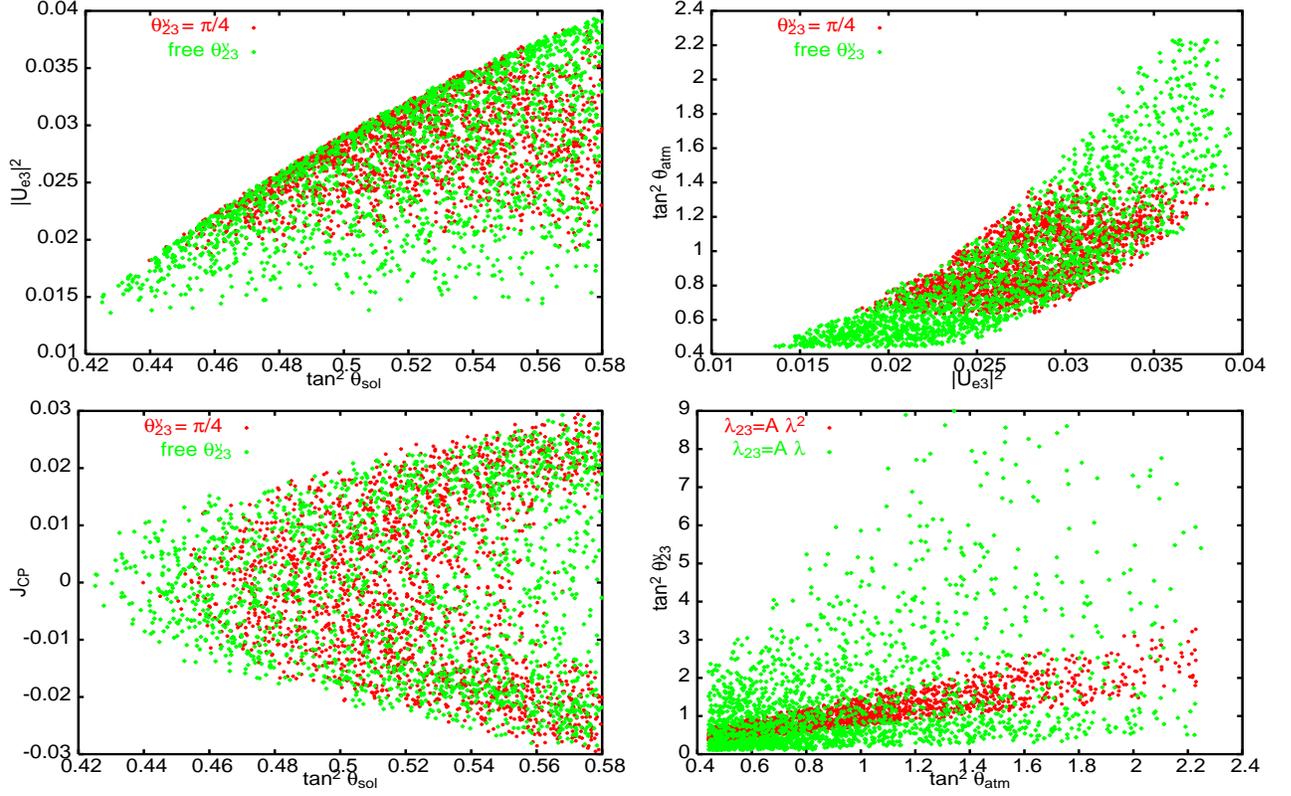,width=19cm,height=24cm}
\vspace{-10cm}
\caption{\label{fig:free/pi4}
Scatter plot of \ts{} against $|U_{e3}|^2$, 
$|U_{e3}|^2$ against $\tan^2 \theta_{\rm atm}$, 
\ts{} against $J_{CP}$ and $\tan^2 \theta_{\rm atm}$
against $\tan^2 \theta_{23}$. 
We choose $\lambda_{12} \equiv \lambda = 0.24$, $\lambda_{13} = \lambda^3$ 
and $\lambda_{23} = \lambda^2$ (except for the lower right plot, 
in which results for $\lambda_{23} = \lambda$ are also shown).  
The neutrino mixing parameters are required
to lie within their respective 3$\sigma$ allowed ranges
of values, given in Eq.\ (\ref{eq:range}).}
\end{center}
\end{figure}

 Let us for illustration fix  
$\lambda_{12} \equiv \lambda = 0.24$  
and choose $\lambda_{13} = A~\lambda^3$ and 
$\lambda_{23} = B~\lambda$ or $\lambda_{23} = B~\lambda^2$. The 
parameters $A,~B$ are let to vary between $1/\sqrt{3}$ 
and $\sqrt{3}$. The 
result of the numerical analysis is displayed in 
Fig.\ \ref{fig:free/pi4}. 
It is seen again from the plots  
that in the case of  ``free'' $\theta_{23}^\nu$ and 
$\lambda_{23} = B~\lambda^2$, 
the oscillation parameters can have
significantly lower values within their 
allowed ranges. 
The reason is that for $\lambda_{23} = \lambda$ there 
are more terms of the same order of 
$\lambda$ contributing to the 
observables. Furthermore, as can be 
seen from the lower right plot, when 
$\theta_{23}^\nu < \pi/4$
($\theta_{23}^\nu > \pi/4$) 
and $\lambda_{23} = \lambda^2$, 
the atmospheric neutrino mixing ``prefers'' 
to take a value $\theta_{\rm atm} < \pi/4$
($\theta_{23}> \pi/4$), 
as discussed above. 

 
\section{\label{sec:real}Realization of the Neutrino Mass Matrix 
within the See--Saw Mechanism}

We will present in this Section 
a very simple see--saw realization of the 
neutrino mass matrix of Eq.\ (\ref{eq:mnul3l2l1}), including
the requisite small perturbation. 
The problem of getting a neutrino mass matrix corresponding to an 
inverted hierarchy within the see--saw mechanism \cite{seesaw} 
has been considered, e.g., in \cite{seesawinv,CAlbright04}. 
The Lagrangian leading to the see--saw mechanism \cite{seesaw} reads 
\be \label{eq:L}
{\cal L} =  
 \frac{1}{2}~(N_R^T)~C^{-1}~M_R~N_R  - 
\overline{N_R}~m_D~L   
+ h.c.,
\ee
%
where $N_R$ are the heavy Majorana neutrinos and $L$ the left--handed  
$SU(2)_L$ lepton doublet. 
The most simple way to generate the requisite 
light neutrino mass matrix $m_\nu$ is to assume 
that the Dirac mass matrix $m_D$ is diagonal 
and that the inverse of the heavy Majorana mass matrix $M_R$ has the 
structure of $m_\nu$ we would like to get.
In this spirit, we take 
\be \label{eq:sperea}
m_D = 
\left(
\bad 
m_{D1} & 0 & 0 \\[0.2cm]
0 & m_{D2} & 0 \\[0.2cm]
0 & 0 & m_{D3} 
\ea 
\right) \mbox{ and } 
M_R^{-1} = 
\left(
\bad 
0 & b & d \\[0.2cm]
\cdot & a & 0 \\[0.2cm]
\cdot & \cdot & 0
\ea 
\right)~,
\ee
where $a, b, d$ are complex, 
i.e., $a = |a| e^{i \alpha}$, $b = |b| e^{i \beta}$ and 
$d = |d| e^{i \gamma}$. 
The resulting light neutrino mass matrix reads 
\be 
m_\nu = - 
\left(
\bad 
0 & b \, m_{D1} \, m_{D2} &  d \, m_{D1} \, m_{D3} \\[0.2cm]
\cdot & a \, m_{D2}^2 & 0 \\[0.2cm]
\cdot & \cdot & 0
\ea 
\right)~,
\ee
which can be identified with the mass 
matrix (\ref{eq:mnul3l2l1}) with a small 
perturbation in the $\mu\mu$ entry. 
In order to match the required ratio of the entries, the 
following conditions must hold:
\be \label{eq:cond}
\bad  
m_{e \mu} \sim m_{e \tau} & \Rightarrow & \frac{\D |b|}{\D |d|} \sim 
\frac{\D m_{D3}}{\D m_{D2} } ~,\\[0.2cm]
m_{\mu \mu} \simeq \epsilon^2 \, m_{e \mu }& \Rightarrow & 
 \epsilon^2 \, \frac{\D m_{D2}}{\D m_{D1} } \simeq \frac{\D |a|}{\D |b|}~.
\ea
\ee
In case of a hierarchy between the Dirac masses, $m_{D3} \gg m_{D2} 
\gg m_{D1}$, we would find that $|b| \gg |d|$ and 
$|b| \gg |a|$ must hold. The ratio between $|d|$ and $|a|$ required to 
reproduce the desired structure of $m_\nu$ would then depend on the ratio 
of the Dirac masses. Looking at Table \ref{tab:break1}, we can identify 
$\tan \theta_{23}^\nu \simeq |d|/|b| \, m_{D3}/m_{D2}$,  
$R \simeq 2 |a\,b^2| \, m_{D2}^4/m_{D1}$ and 
$m_3  \simeq |a\,d| \, m_{D2}^2 \, m_{D3}$.  
The phase $\phi$, on which $J_{CP}$ 
depends in the case of
hierarchical charged lepton mixing, coincides with $\alpha - \beta$, 
see Section \ref{sec:CPVgen}. 
Note that although the structure of $M_R^{-1}$ is identical to that of 
$m_\nu$, the 12 element of $M_R^{-1}$ is, for hierarchical Dirac masses, 
much larger than the other two non--vanishing elements. 

   In analogy to the procedure outlined above in Section \ref{sec:CPVgen}, 
we can simplify the diagonalization of $M_R^{-1}$ by writing it as 
\bea 
M_R^{-1} = {\rm diag}( e^{i(\beta - \alpha/2)}, e^{i \alpha/2}, 
e^{i(\alpha/2 + \gamma - \beta )}) \, 
\left(
\bad 
0 & |b| & |d| \\[0.2cm]
\cdot & |a| & 0 \\[0.2cm]
\cdot & \cdot & 0
\ea 
\right) \, 
{\rm diag}( e^{i(\beta - \alpha/2)}, e^{i \alpha/2}, 
e^{i(\alpha/2 + \gamma - \beta )}) \\[0.3cm]
\equiv P \, 
\left(
\bad 
0 & |b| & |d| \\[0.2cm]
\cdot & |a| & 0 \\[0.2cm]
\cdot & \cdot & 0
\ea 
\right) \, P \equiv  P \, \left|  M_R^{-1} \right| \, P ~,
\eea
%
so that we do not have to bother about the phases in the 
procedure. Diagonalizing 
$|M_R^{-1}|$ via $V_R \, |M_R^{-1}| \, V_R^T = 1/|M_R^{\rm diag}|$  
brings the Majorana neutrinos in their physical mass basis, in which we 
have to replace $m_D$ with $\tilde{m}_D \equiv V_R~P~m_D$. 
Defining $\epsilon_1 = |d/b|$ and $\epsilon_2 = |a/b|$, 
$V_R$ is approximately given by 
\be 
V_R \simeq 
\left( 
\bad 
\sqrt{\frac{1}{2}} + \frac{\epsilon_2}{4 \sqrt{2}} & 
-\sqrt{\frac{1}{2}} + \frac{\epsilon_2}{4 \sqrt{2}} & 
- \frac{\epsilon_1}{\sqrt{2}} \\[0.3cm] 
\sqrt{\frac{1}{2}} - \frac{\epsilon_2}{4 \sqrt{2}} & 
-\sqrt{\frac{1}{2}} + \frac{\epsilon_2}{4 \sqrt{2}} & 
 \frac{\epsilon_1}{\sqrt{2}} \\[0.3cm] 
\epsilon_2 \, \epsilon_1  & -\epsilon_1 & 1 
\ea 
\right) + {\cal{O}}(\epsilon^2)~,
\ee
with the corresponding eigenvalues $\pm |b| \, (1 \mp \epsilon_2/2) 
= |a|/2 \pm |b|$, and $|a \, d^2/b^2|$.\\ 
%
 
 Consider now leptogenesis  \cite{lepto} in the 
scenario under study. Decays of heavy Majorana neutrinos in the early 
Universe create a lepton asymmetry which is subsequently converted 
into a baryon asymmetry (for reviews see, e.g., \cite{leptorev}). 
The requisite $CP$--violating asymmetry 
is caused by the interference of 
the tree level contribution and 
the one--loop corrections in the decay rate
of the three heavy 
Majorana neutrinos, 
$N_1 \ra \Phi^- \, \ell^+$ and $N_1 \ra \Phi^+ \, \ell^-$ and 
it reads \cite{leptorev}: 
\begin{equation} \label{eq:eps}
\ba 
\varepsilon_i = \frac{\D \Gamma (N_i \ra \Phi^- \, \ell^+) - 
\Gamma (N_i \ra \Phi^+ \, \ell^-)}{\D \Gamma (N_i \ra \Phi^- \, \ell^+) +  
\Gamma (N_i \ra \Phi^+ \, \ell^-)} \\[0.4cm]
\simeq  \D \frac{\D 1}{\D 8 \, \pi \, v^2} \frac{\D 1}
{(\tilde{m}_D \tilde{m}_D^\dagger)_{ii}} 
\sum_{j=2,3} {\rm Im} (\tilde{m}_D \tilde{m}_D^\dagger)^2_{ij} 
\, f(M_j^2/M_i^2) ~,
\ea 
\end{equation}
where $\Phi$ is the Higgs field, $v$ its vacuum expectation value and  
$f(x) \simeq -3/(2 \sqrt{x})$ for $x \gg 1$. Explicitly, the 
function reads 
\be \label{eq:f} 
f(x) = \sqrt{x} 
\left( 
1 - (1 + x) \log (1 + 1/x) + \frac{1}{1 - x} 
\right)~.
\ee
From the above expression for $M_R^{-1}$ the right--handed Majorana 
masses are found to be 
\be \label{eq:Mspre}
M_{1,2} \simeq \pm \frac{1}{|b|} \, (1 \pm \epsilon_2 /2)
\mbox{ and } M_3 \simeq \left| \frac{b^2}{a\, d^2} \right| ~,
\ee
displaying an ``inverted--hierarchy--like'' spectrum, 
where now however the two lighter neutrinos are close in mass with 
a very small mass difference of $\Delta M \simeq |a/b^2|$. 
This can lead to a resonant amplification of the decay asymmetry 
\cite{leptores}. The relevant condition under which the formula 
Eq.\ (\ref{eq:eps}) is valid (otherwise a more complicated formula 
is required) can be written as 
$\Delta M \gg \Gamma_i$, where $\Delta M$ is the mass difference of the 
two neutrinos and $\Gamma_i$ the tree level decay width, which is 
$\Gamma_i = M_i \, (\tilde{m}_D \tilde{m}_D^\dagger)_{ii}/(16 \pi \, v^2)$. 
In our case we find that 
\be
\frac{\Delta M}{\Gamma_1} \simeq \frac{\Delta M}{\Gamma_2} 
\simeq \frac{32 \pi \, v^2}{m_{D2}^2} \, \epsilon_2~.
\ee 
To simplify the analysis, we can choose now 
$m_{D2} \simeq \lambda^n \, m_{D3} $ and 
$m_{D1} \simeq \lambda^m \, m_{D3} $, where the expansion parameter 
$\lambda \simeq 0.23$. 
To reproduce the mass matrix (\ref{eq:mnu2}) 
for the entries in $M_R^{-1}$ it must hold: 
$|d| \simeq |b| \, \lambda^n$ and $|a| \simeq |b| \, \lambda^l$, 
with $l = 2 + m - n$, or $\epsilon_2 = \lambda^l$.  
For the lightest neutrino $M_1$ we have explicitly 
\bea
\varepsilon_1 \simeq \frac{\D 1}{\D 8 \pi \, v^2} \, 
\frac{\D 1}
{\D (\tilde{m}_D \tilde{m}_D^\dagger)_{11}} \, 
\left( 
{\rm Im} (\tilde{m}_D \tilde{m}_D^\dagger)^2_{12} 
\, f(M_2^2/M_1^2) + 
{\rm Im} (\tilde{m}_D \tilde{m}_D^\dagger)^2_{13} 
\, f(M_3^2/M_1^2)
\right)~.
\eea
We can simplify the function $f$ in the two terms with 
$ f(M_2^2/M_1^2) \simeq  M_1^2/(M_1^2 - M_2^2) 
\simeq -1/2 \, \lambda^{-l}$ 
for the two neutrinos close in mass and 
$ f(M_3^2/M_1^2) \simeq -3/2 \, M_1/M_3 \simeq -3/2 \, \lambda^{l + 2n}$. 
After straightforward calculation, putting everything together yields:    
\be 
\varepsilon_1 \simeq \frac{-1}{32 \pi} \, \frac{m_{D3}^2}{v^2} \, 
\frac{1}{\lambda^{2 m} + 2 \lambda^{2 n}} \, 
\left\{ 
\frac{\lambda^{2 m} - 2 \lambda^{2 n}}{\lambda^{l}} \, \sin 2 (\alpha-\beta) 
+ 6 \, \lambda^{l + 2n} \, \left(\lambda^{3 n} + \lambda^{l+2m+n}
\right)^2 
\right\}~.
\ee
Note that the first term is proportional to 
$\sin 2 (\alpha-\beta) = \sin 2 \phi$. 

  The important parameter governing the wash--out \cite{leptorev} 
of the generated lepton asymmetry is 
\be 
\tilde{m}_1 \simeq \frac{\sqrt{\dma}}{2} \, \left( 
\lambda^{m-n} + 2 \lambda^{n-m}
\right)~,
\ee
%
where we have used that $m_{D3}^2 \, |b| \, \lambda^{m+n} \simeq \sqrt{\dma}$, 
which is obvious from Eqs.\ (\ref{eq:mnul3l2l1})
and (\ref{eq:cond}). 
A typical choice of parameters leading to reasonable heavy Majorana 
masses is $m_{D3} \simeq v$ and $m = n = 1$, which leads to 
$\Delta M/\Gamma_1 \simeq 96 \pi$, $M_{1,2} \simeq 3 \cdot 10^{13}$ GeV 
and $M_3 \simeq 10^{16}$ GeV. 
The implied heavy Majorana mass matrix is 
\be 
M_R \simeq \frac{1}{|b| \, \lambda^4} \, 
\left( 
\bad 
0 & 0 & \lambda^3 \\[0.2cm]
\cdot & \lambda^2 & \lambda \\[0.2cm]
\cdot & \cdot & 1
\ea
\right)~.
\ee
The decay asymmetry is dominated by the first term and given by 
\be 
\varepsilon_1 \simeq \frac{1}{96 \pi \, \lambda^2} \, 
\sin 2 \phi \simeq 6.3 \cdot 10^{-2} \, \sin 2 \phi~.
\ee
Hence, the decay asymmetry is (due to the two Majorana neutrinos 
close in mass) rather large and depends to leading order 
on two times the Dirac phase. Since typically negative and small 
$|\varepsilon_1| \gs 10^{-5}$  
is required for successful leptogenesis \cite{leptorev}, one expects 
a phase closely below $\pi$ or $2\pi$ 
and therefore small $CP$--violating effects in 
neutrino oscillations. Since $\cos \phi$ has to be close to +1 in order 
to obey the constraints coming from the correlations between 
$|U_{e3}|$ and \ts
expressed in Eq.\ (\ref{eq:corCPV}), 
 $\phi$ consistent with all constraints
should have a value close to, but somewhat smaller than,
$2\pi$.


\section{\label{sec:concl}Conclusions}


We analyzed the possibility that the observed pattern of neutrino mixing 
in the PMNS matrix arises as a result of an interplay 
between  the two unitary matrices 
associated with the diagonalization of the charged lepton ($m_\ell$) 
and neutrino ($m_\nu$) mass matrices, $U_{\rm PMNS} = U_L^\dagger U_\nu$. 
The matrix $U_\nu$ diagonalizing the neutrino Majorana mass matrix $m_\nu$, 
is first assumed to have an exact bimaximal mixing form. 
We have summarized 
the allowed structures of the charged lepton 
mixing matrix $U_L$ in this case and have given 
the implied forms of the matrices $m_\ell m_\ell^\dagger$ and 
$m_\ell = m_\ell^T$, diagonalized by $U_L$.  
We have assumed further that the 
origin of bimaximal mixing is a weakly  
broken flavor symmetry  
corresponding to the conservation of the non--standard lepton charge
$L' = L_e - L_\mu - L_\tau$.   
The latter does not predict, in general, the mixing angle 
$\theta_{23}^\nu$ in $U_\nu$ to be maximal. 
We therefore generalized our analysis to the case 
of an a priori non--maximal mixing angle $\theta_{23}^\nu$. 
We performed a detailed study of the 
allowed structures of the charged lepton mixing 
and the implied correlations between 
the neutrino mixing and $CP$--violating 
observables. Considering the case when
the $L_e - L_\mu - L_\tau$ flavor symmetry
is broken by one small element in $m_{\nu}$,
we found that the three $CP$--violating 
rephasing invariants, associated with
the three $CP$--violating phases in the PMNS mixing matrix,
are related. This result is valid 
independently of the location of the
symmetry breaking element in $m_\nu$. 
We discussed the corrections 
from hierarchical charged lepton mixing and their 
impact on the precise value of $\theta_{\rm atm}$.
To leading order,  
$\theta_{23}^\nu$ corresponds to $\theta_{\rm atm}$.
The Dirac phase $\phi$, measurable in 
neutrino oscillation experiments, 
enters into the expression for $\ts$:
$\ts \simeq 1 - 4\, \cos \phi~ \cot \theta_{\rm atm}~|U_{e3}|$. 
Since $\ts$ is constrained from above 
by the data, $\ts \ls 0.58$, 
this result implies that the smaller $|U_{e3}|$,
the closer to 0 or $2\pi$ the phase $\phi$ would be.
Consequently, for $|U_{e3}| \ls 0.1$, 
$J_{CP}$ would be additionally suppressed 
by the fact that $\sin \phi$ has to be 
relatively small. Moreover, $J_{CP}$ is also suppressed 
when $m_\nu$ conserves $CP$. 
The possible 
``tension'' between a relatively 
small value of 
$\theta_{13}$ and substantially 
non--maximal solar neutrino mixing 
is somewhat relaxed in the case of 
$\theta_{23}^\nu \neq \pi/4$. 
More specifically, the lower limit on 
$|U_{e3}|$ obtained for exact bimaximal mixing 
can be reduced by a factor of two, 
i.e., to $|U_{e3}^2| \gs 0.007$. 
Moreover, in contrast to the case 
of $\theta_{23}^\nu = \pi/4$, $\sin^2 2\theta_{\rm atm}$ 
can take any value inside its currently 
allowed range.  However, if  $\theta_{\rm atm}$ is 
found to be very close to $\pi/4$, the limits 
corresponding to the case of exact bimaximal mixing are of course 
recovered. As a consequence of the obtained 
correlations between the neutrino 
mixing and $CP$--violating observables,
the effective Majorana 
mass in \onbb{} depends on $J_{CP}$ 
and $\sin^2 \theta_{\rm atm}$.
A very simple see--saw realization 
of a neutrino mass matrix conserving $L_e - L_\mu - L_\tau$ was 
written down, which contains one heavy Majorana neutrino with mass 
much larger than the other two, which in turn are close in mass. 
In this scheme, the Dirac phase responsible for $CP$ violating effects 
in neutrino oscillations is also responsible for 
the baryon asymmetry of the Universe. It has to be closely below $2\pi$, 
which is consistent with the mentioned relation between 
this phase, $\ts$ and $|U_{e3}|$.

\vspace{0.5cm}
\begin{center}
{\bf Acknowledgments}
\end{center} 
This work was supported by the EC network HPRN-CT-2000-00152 (W.R.) 
and by the Italian INFN under the program ``Fisica Astroparticellare''
(S.T.P.).

\pagestyle{empty}

\end{document}